# Revealing a Systematic High-latitude Current Sheet at Jupiter

Yan Xu[1,2], Zhonghua Yao[2]*, Frederic Allegrini[3,4], Shengyi Ye[1]*, Binzheng Zhang[2]*, Zhili Zeng[2], Enhao Feng[5], Jiuwen Sun[6], William Dunn[7], Scott Bolton[3]

[1] Department of Earth and Space Sciences, Southern University of Science and Technology (SUSTech), Shenzhen, China

[2] Department of Earth Sciences, the University of Hong Kong, Pokfulam, Hong Kong, SAR, China

[3] Southwest Research Institute, San Antonio, Texas, USA.

[4] Department of Physics and Astronomy, University of Texas at San Antonio, San Antonio, Texas, USA.

[5] School of Space and Earth Sciences, Beihang University, Beijing, China.

[6] School of Earth and Space Sciences, Peking University, Beijing, China.

[7] Mullard Space Science Laboratory, University College London, Dorking, UK.

Correspondence and requests for materials should be addressed to Zhonghua Yao (yaozh@hku.hk), Shengyi Ye (yesy@sustech.edu.cn) and Binzheng Zhang (binzh@hku.hk)

**Abstract**

Based on models derived from Earth's magnetotail, other planets with dipole magnetic fields, including Mercury, Jupiter, and Saturn, were expected to possess similar magnetotail configurations. In this traditional picture, the majority of plasma is confined near the magnetic equator within a plasma sheet (or plasma disc), whereas higher-latitude regions feature strong magnetic fields that are open to the solar wind, forming magnetospheric lobes. However, auroral observations and recent simulations have shown that Jupiter's magnetic topology differs markedly from this picture, particularly in its high-latitude regions where magnetic field lines are predominantly closed. This discrepancy calls for a re-examination of high-latitude magnetospheric structure at Jupiter. Here, using Juno measurements acquired between 2016 and 2022, we show that Jupiter's nightside high latitudes host a persistent current-sheet-like structure above ~40° magnetic latitude near midnight. This structure contains internally sourced oxygen and sulfur ions and exhibits azimuthal magnetic signatures opposite to the equatorial current sheet's bend-back. These findings indicate that the canonical picture of planetary magnetotail architecture requires revision. Our results provide new insight into the architecture of rapidly rotating magnetospheres and offer a framework for interpreting magnetospheric structures at exoplanets.

**Introduction**

Global dipolar magnetic fields are common among solar system planets, and these fields are continuously compressed by the high-speed solar wind, forming planetary magnetospheres with long nightside magnetotails[1,8]. Decades of terrestrial observations have established the canonical structure of such magnetotails: a thin equatorial current sheet, where oppositely directed magnetic fields compress plasma into a confined layer, separates expansive, plasma-poor magnetic lobes that remain open to the solar wind[9–11]. These lobes map to the polar ionosphere, forming Earth's polar caps, where the resulting low particle flux produces dim auroral emissions[12,13]. Compared to Earth, Jupiter's rapid 10-hour rotation, together with the substantial heavy-ion outflow from Io (~1,000 kg s$^{-1}$), creates a rotationally dominated system in which the middle and outer magnetosphere form a thick magnetodisc supported by centrifugal forces and plasma pressure[14–16]. Despite these fundamental differences, Jupiter's high-latitude nightside magnetosphere had traditionally been described using Earth-like frameworks, invoking a central plasma sheet flanked by plasma-depleted lobes on open field lines[1–3].

Remote sensing auroral observations have challenged this conventional picture. Unlike Earth's polar cap—where open magnetic field lines connect to interplanetary space and contain depleted plasma[17]—Jupiter's polar regions exhibit persistently bright ultraviolet aurora[18,19] that remain active even under weak solar wind driving[20,21], indicating a fundamentally different magnetic topology. These quasi-steady emissions, unlike Earth's sporadic polar rain aurora [22], imply that Jupiter's high-latitude regions are largely threaded by closed field lines [5,6,23] connecting both hemispheres through the outer magnetosphere rather than extending into interplanetary space. In situ measurements from Juno at 6–7 $R_J$ support this interpretation: the spacecraft encountered closed polar field lines carrying internally sourced plasma with heavy-ion signatures (e.g., $O^+$, $S^{2+}$) characteristic of the magnetotail [24], indicating closed magnetic linkage to Jupiter's magnetosphere rather than open to the solar wind, and high-density magnetospheric features in the pre-midnight high-latitude region have been documented adjacent to the cusp region associated with open field lines[25]. Additionally, evidence for open magnetic flux has been revealed by anticorotational flows and helium ions in Jupiter's mid-latitudes[26]. Global MHD simulations further predict that these closed polar flux tubes extend into the nightside magnetosphere in helical structures due to rapid rotation[5–7], though their global features remain observationally unconstrained.

This evidence for closed polar flux tubes raises a fundamental question: where do these field lines extend to in the magnetosphere, and what signatures does this structure produce? If closed field lines threading Jupiter's polar regions map into the nightside magnetosphere, they would be expected to be associated with a plasma population at high latitudes, rather than with the plasma-depleted open-lobe configuration familiar at

Earth. Yet no such structure has been comprehensively investigated in Jupiter's middle-to-far magnetosphere. Past flyby missions, including Ulysses and New Horizons, only provided brief views of Jupiter's low-latitude magnetosphere or single-track samplings, without resolving the three-dimensional structure of the high-latitude closed flux tubes that map to the polar caps[14,27]. Galileo, the only long-term in situ survey prior to Juno, remained near the equatorial plasma sheet, so that the high-latitude tail lobes above the main auroral oval were left essentially unexplored[28,29]. Juno's polar, highly eccentric orbit now provides unprecedented, repeated access to the polar magnetosphere and high-latitude nightside closed field lines[30,31].

This study uses Juno datasets from 2016 to 2022 that traverse the full nightside magnetosphere, combining comprehensive fields[32] and particle measurements[33] with high-resolution global MHD simulations[5,34]. We identify a previously unrecognized high-latitude dense structure above ~40° magnetic latitude in the midnight sector that carries internally sourced heavy ions ($O^+$, $S^{2+}$) with similar populations to those in the magnetotail current sheet and exhibits reversed azimuthal magnetic signatures opposite to the equatorial sheet's bend-back configuration. These findings provide the first observational evidence of a surprising high-latitude current sheet-like structure.

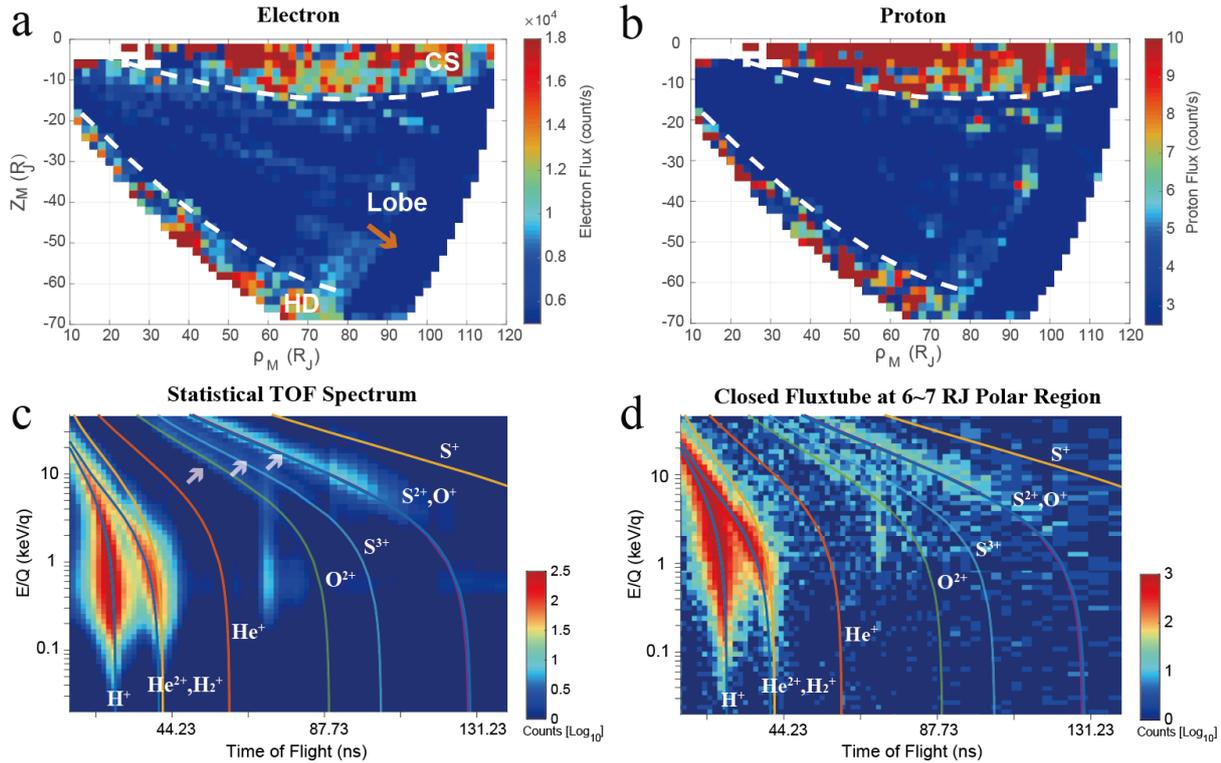

**Figure 1. Global observational distributions of particle fluxes at Jupiter and low-energy particle composition analysis in high-latitude dense regions.** Statistical maps of **(a)** total electron flux (100–1000 eV) and **(b)** total proton flux (1–50 keV) in magnetic coordinates ($Z_M$, $\rho_M$ in $R_J$), where $\rho_M = \sqrt{X_M^2 + Y_M^2}$. The Z–$\rho_M$ domain is divided into 2 $R_J$ × 2 $R_J$ grid cells, each colored by the median flux value when at least three data points are available. White dashed curves in panels a and b delineate the

boundaries separating the low-latitude current sheet, the lobe region, and the high-latitude dense region, derived from sinusoidal fits: $Y_1=14.85\times\sin(0.02079X_1+3.042)$ and $Y_2=64.27\times\sin(0.0156X_2+3.227)$, consistent with the partitioning used in Fig. 3a and 3c; **(c)** Statistical time-of-flight (TOF) spectrum for the high-latitude dense region, defined as locations where magnetic latitude exceeds 35° and the summed electron flux (100–1000 eV) is greater than 8000 counts s$^{-1}$. The grey arrow highlights the enhanced internally sourced oxygen and sulfur ions in Jupiter's magnetosphere; **(d)** TOF spectrum from a closed flux tube crossing at 6–7 $R_J$ near the southern polar region on September 3, 2021, reported by Szalay et al.[24], showing enhanced heavy-ion signatures (O$^+$, S$^{2+}$) of internal magnetospheric origin. Panel (c) exhibits distributions of magnetospheric O$^+$ and S$^{2+}$ consistent with panel (d), indicating that the high-latitude dense region contains plasma of similar magnetospheric origin.

**Results**

**Global Distribution of Particles and Identification of High-Latitude Dense Region.**
We constructed global statistical maps of electron and proton fluxes across Jupiter's nightside magnetosphere using Juno JADE measurements from Perijoves 1–41 (2016–2022). Fig. 1a and 1b reveal three spatially distinct regions in the nightside sector (18–06 LT). The first is the low-latitude current sheet (CS), characterized by peak particle fluxes near the magnetic equator (Z ≈ 0) and extending radially from roughly 30 to beyond 100 $R_J$. The second is the mid-latitude lobe, which exhibits significantly depleted fluxes at intermediate magnetic latitudes (approximately 20° < |θ| < 40°), which is consistent with the open flux regions explored by Delamere et al.[26]. The third is a previously unrecognized high-latitude dense (HD) region, marked by enhanced particle fluxes at magnetic latitudes above about 40° and heights |Z| > 50 $R_J$. This HD region lies southward of the conventional lobe and forms a persistent dense plasma structure throughout the nightside magnetosphere. Because Juno's perijove gradually shifts to higher latitudes as it moves from the morning side into the nightside and duskside sectors[30], the HD region is most frequently observed in the duskside and pre-midnight sectors, where Juno spends most of its time in the high-latitude magnetosphere. The distributions of particle fluxes in different local time sectors are shown in Supplementary Fig. 1. The white dashed curves in panels (a) and (b) delineate the fitting boundaries separating the low-latitude current sheet, the lobe region, and the high-latitude dense region, enabling systematic statistical analysis of magnetic field characteristics across distinct magnetospheric domains. The potential effects of Juno's orbital factors on the statistical results are shown in Supplementary Fig. 2.

The compositional characteristics of the HD region provide strong evidence for its magnetospheric origin. Fig. 1c presents the time-of-flight (TOF) spectrum of low-energy plasma, showing that the HD region contains substantial heavy-ion populations (e.g., O$^+$, S$^{2+}$) similar to those in the equatorial CS and clearly distinct from the solar-wind–dominated composition expected along open field lines (statistics details of Fig. 1c in Methods). The superposed epoch electron energy spectrum at different heights from the magnetic equator is presented in Supplementary Fig. 3. Both an enhanced

current-sheet–like electron spectrum at 500–3000 eV and an enhanced magnetosheath-like spectrum below 100 eV appear simultaneously at high latitudes. This indicates that, in the high-latitude dusk sector, the potential cusp event indicated by magnetosheath-like features[25] and the HD event both can be observed, consistent with the case shown in Fig. 2. Moreover, the HD region, which is located far from the plasma disc and has traditionally been classified as part of the lobe connected to open magnetic field lines[31], exhibits a composition that closely matches that of the closed-flux region near the southern pole identified by Szalay et al.[24] (Fig. 1d), as well as the plasma composition observed in the distant magnetotail current sheet. This consistency indicates that the dense plasma in the HD region may form part of a continuous pathway linking closed magnetic flux from the polar region through the nightside magnetotail and potentially back to the opposite hemisphere.

**Case Study: High-Latitude Dense Region Crossing on April 20, 2022.** To illustrate the characteristic signatures of the HD region, we examine a representative nightside crossing on April 20, 2022 (orbit PJ 41). As recently analyzed by Xu et al.[25], during April 14, 2022 Juno was understood to travel into the cusp and dense regions (including their mixed transition zone) in the pre-midnight sector. Throughout the crossing on April 20, 2022 in this study (Fig. 2), Juno remained in very high latitude (Fig. 2h) and the magnetic field consistently exhibited a directional reversal relative to the bend-back configuration ($B_\varphi < 0$). The particle measurements provide additional evidence for the distinct nature of the HD region. Fig. 2d–f show electron and ion energy spectrograms during the crossing. The polar cap/lobe region displays severely depleted particle fluxes across all energies, consistent with expectations. In contrast, the HD region exhibits enhanced fluxes of electrons in the ~0.1–1 keV range, protons in the ~0.4–30 keV range, and heavy ions in the ~0.3–50 keV range. These spectra closely resemble those of the equatorial CS rather than the depleted lobe, which is unexpected at such high magnetic latitudes. Superposed epoch electron energy spectrum at different heights from the magnetic equator is shown in Supplementary Fig. 3. The spacecraft's magnetic footprint trajectory (Supplementary Fig. 4g, h) shows that during the HD interval, Juno's footprint mapped to latitudes poleward of the main auroral oval (white dashed curve), within the polar emission region, confirming the high-latitude nature of this dense plasma domain.

Quantitative comparison of electron energy distributions further underscores the similarity between the HD region and the equatorial CS. Fig. 2h presents the averaged electron flux as a function of energy for four regions: the magnetosheath (red), CS (orange), magnetosheath-like region (green), and HD region (blue). The magnetosheath shows elevated electron fluxes near ~100 eV that drop rapidly with increasing energy. The magnetosheath-like region displays a low-energy–dominated spectrum like magnetosheath, which could be linked to dusk-side cusp region as identified in Xu et al[25,35]. In contrast, both the CS and HD regions exhibit comparable flux levels of ~$10^3$–$10^4$ counts s$^{-1}$ near 1 keV and share similar spectral slopes. Although their peak energies differ slightly—approximately 1 keV for the HD region and 3 keV for the CS—the overall flux shapes remain similar. This similarity indicates that the HD region contains magnetospheric plasma with density properties comparable to those of the CS. The HD energy distribution shows no evidence of the low-energy magnetosheath-like electron characteristic related to open field or boundary-layer populations.

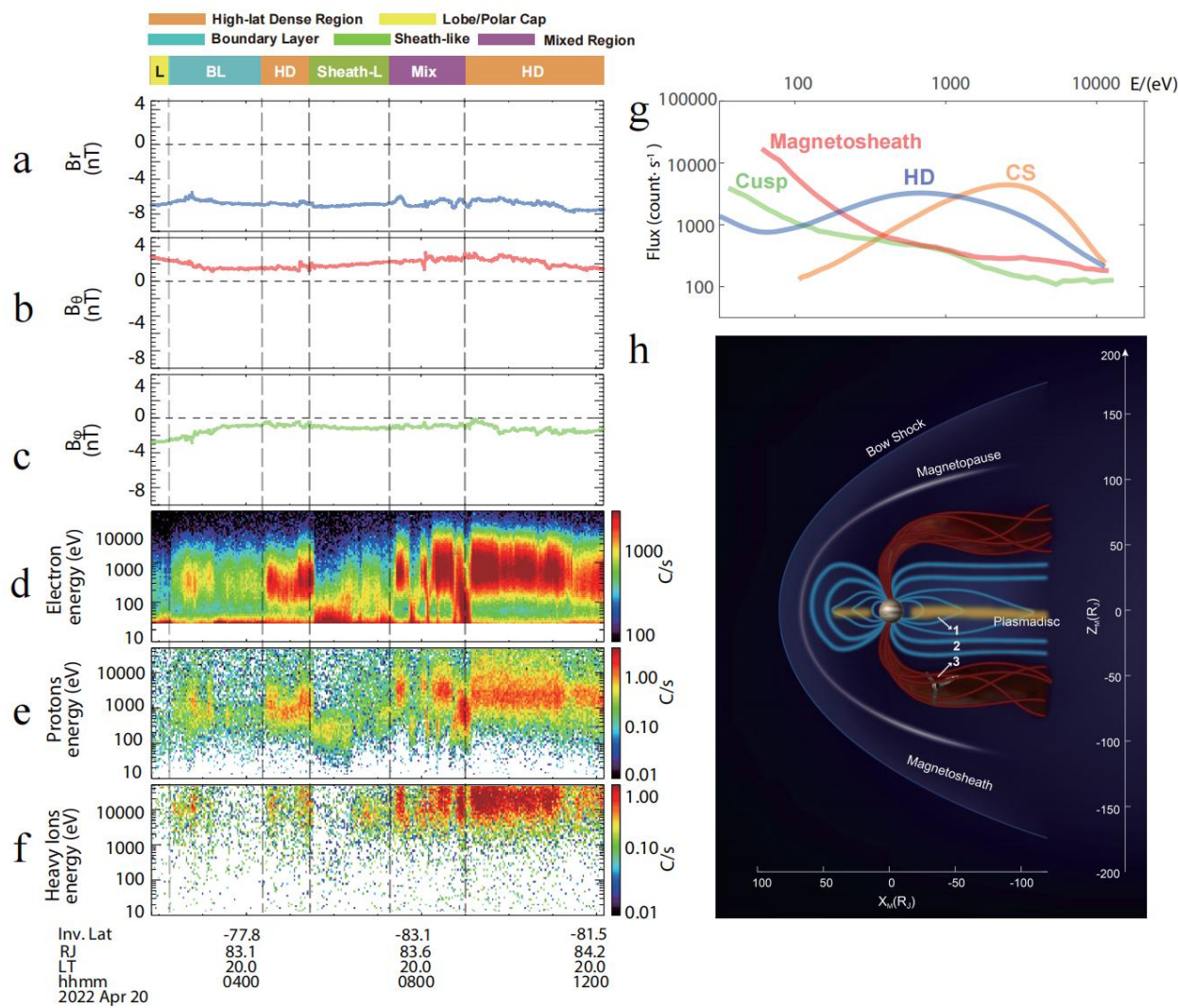

**Figure 2. Juno observations on a HD case in the pre-midnight sector on April 20, 2022.** (**a-c**) Magnetic field components ($B_r$, $B_\theta$, $B_\varphi$) in spherical magnetic coordinates from the Juno MAG instrument during a pre-midnight sector crossing. (**d-f**) Energy–time spectrograms of electrons, protons, and heavy ions from the JADE instrument, with color scales in counts per second. Colored bars at the top mark the identified plasma regions: lobe/polar cap (L), boundary layer (BL), high-latitude dense region (HD), magnetosheath-like region, and mixed region. The bottom axis shows invariable latitude from −77.8° to −83.1°, radial distance, local time and Universal Time (UT) in hhmm format. (**g**) Electron flux–energy distributions for the identified regions: magnetosheath (red), equatorial current sheet (orange), magnetosheath-like region (green), and HD region (blue). The HD region exhibits flux characteristics similar to those of the equatorial current sheet. (**h**) Illustration of the spacecraft location and the HD region in Jupiter's high-latitude magnetosphere as detected by Juno. The spacecraft

icon marks Juno's high-latitude position in magnetic coordinates, where it measured particle distributions characteristic of a current-sheet–like structure. The numbers 1, 2, and 3 point to the CS, lobe, and HD regions, respectively, consistent with those shown in Fig. 1a and Fig. 4a. The magnetopause and bow shock locations shown in panel H are taken from the Joy et al. [36] model for a solar wind dynamic pressure of 0.2 nPa.

**Global MHD Simulation Results.** To visualize the global structure of high-latitude closed flux tubes, we performed global magnetohydrodynamic (MHD) simulations of Jupiter's magnetosphere using the GAMERA model[5,34,37,38] (simulation details in Methods). Fig. 4a shows the plasma number density in a noon–midnight meridional cross-section. The simulation results are consistent with the three-region structure observed by Juno: (1) a dense low-latitude current sheet confined within $|Z| < 20$ $R_J$, (2) a depleted mid-latitude lobe region with depleted densities at heights $20 < |Z| < 50$ $R_J$, and (3) a high-latitude dense region at $|Z| > 50$ $R_J$ where densities increase to values intermediate between those of the lobe and the current sheet. The magnetic field topology in the simulation exhibits the same azimuthal reversal observed in the Juno data. Fig. 4b shows the distribution of the azimuthal magnetic field component $B_\varphi$. The equatorial magnetotail displays positive $B_\varphi$, characteristic of the expected bend-back configuration. The lobe regions at intermediate latitudes also show positive $B_\varphi$, consistent with stretched tail-like flux. In contrast, the nightside high-latitude region ($|Z| > 50$ $R_J$) exhibits a distinct 'reverse spiral' structure with negative $B_\varphi$, closely matching the observational signature in Fig. 3c. The simulation demonstrates that this $B_\varphi$ reversal is a persistent feature of Jupiter's high-latitude magnetosphere.

The three-dimensional magnetic field line topology reveals the spatial organization of high-latitude closed flux (Fig. 4c). Blue field lines represent conventional low-latitude closed flux tubes that originate from equatorial ionospheric latitudes, extend into the nightside magnetotail with the characteristic bend-back geometry, and remain confined below ~30 $R_J$ height. These field lines form the classical magnetodisc/current sheet configuration well-documented in previous studies[15,28]. In contrast, red field lines represent high-latitude closed flux tubes originating from polar ionospheric regions at latitudes poleward of the main auroral oval. These flux tubes rise to heights $|Z| > 40$ $R_J$, extend deep into the distant dawnside magnetotail beyond 100 $R_J$, and eventually return planetward to close at the opposite polar region. Complete views of simulated Jupiter's high-latitude closed magnetic field lines in this study and comparison with Earth's lobe structure are shown in Supplementary Fig. 6-7. The red field lines follow steeply inclined trajectories through the high-latitude magnetosphere. This simulated topology provides a global picture for understanding the HD region as the nightside cross-section of these high-latitude closed flux tubes, which carry magnetospheric plasma between the polar regions through the distant magnetotail.

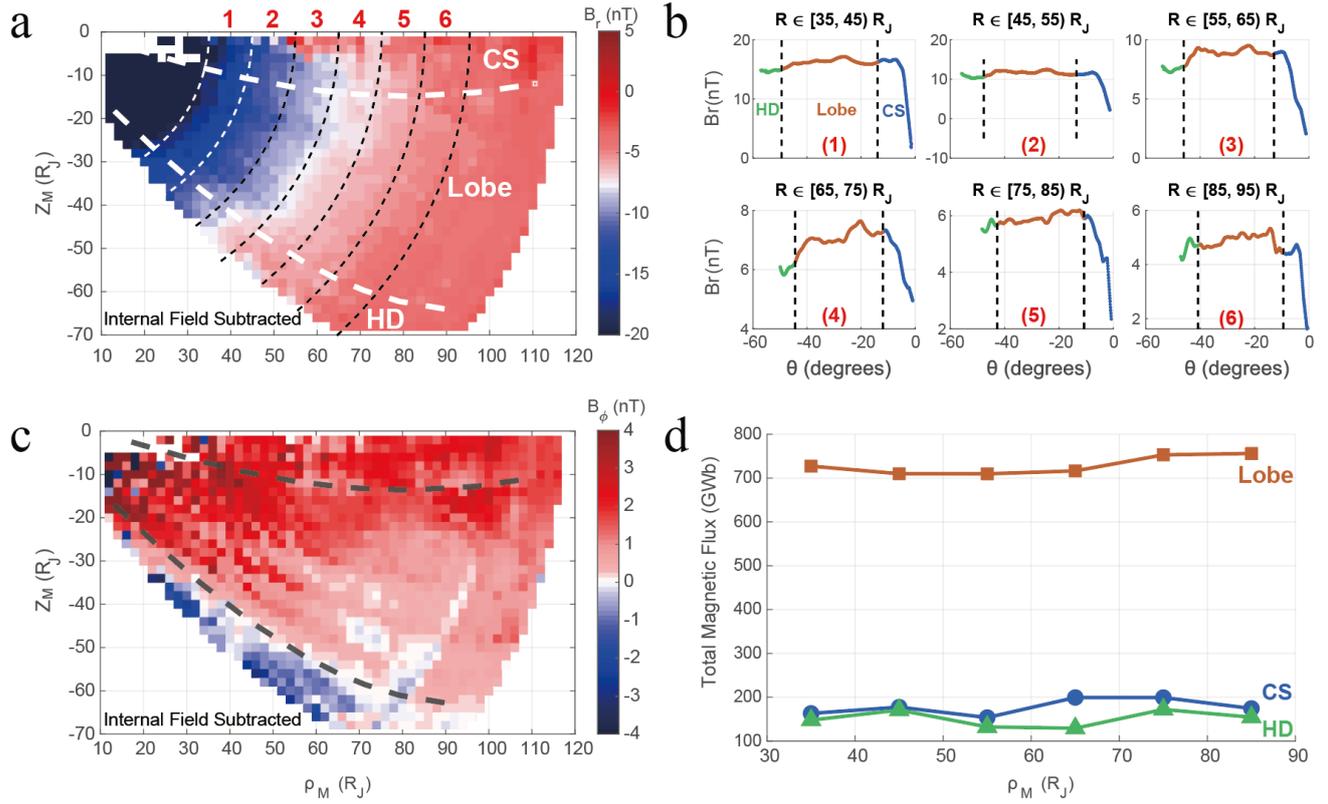

**Figure 3. Statistical magnetic field characteristics of Jupiter's nightside magnetosphere.** Data shown in magnetic coordinates are from the Juno MAG instrument through the same investigation time range and the same statistical method as Fig. 1. (**a**) Radial magnetic field component $B_r$ with the JRM33 internal field[39] removed. White dashed curves mark the boundaries separating the low-latitude current sheet (CS), the lobe, and the HD region, determined from particle flux distributions (Fig. 1a, b). Black arcs indicate six radial bins (35–45, 45–55, 55–65, 65–75, 75–85, and 85–95 $R_J$) used for averaging in panel b, as marked by numbers 1-6. (**b**) Averaged $B_r$ profiles as a function of magnetic latitude θ for the six radial bins, each spanning 10 $R_J$. The HD region (green) maintains weak field strengths comparable to the equatorial CS (blue), in contrast to the strong-field lobe (orange/red). (**c**) Azimuthal magnetic field component $B_\varphi$. The HD region exhibits reversed $B_\varphi$ signatures inconsistent with the bend-back topology characteristic of the lobe. (**d**) Integrated magnetic flux for each region under an assumed axisymmetry. The lobe region contains more than 700 GWb of flux (orange squares), exceeding the estimated magnetic flux in Jupiter's dark polar caps (338 ± 139 GWb)[40]. This flux excess indicates that the lobe region contains a substantial fraction of closed field lines rather than being composed entirely of open flux connected to the solar wind.

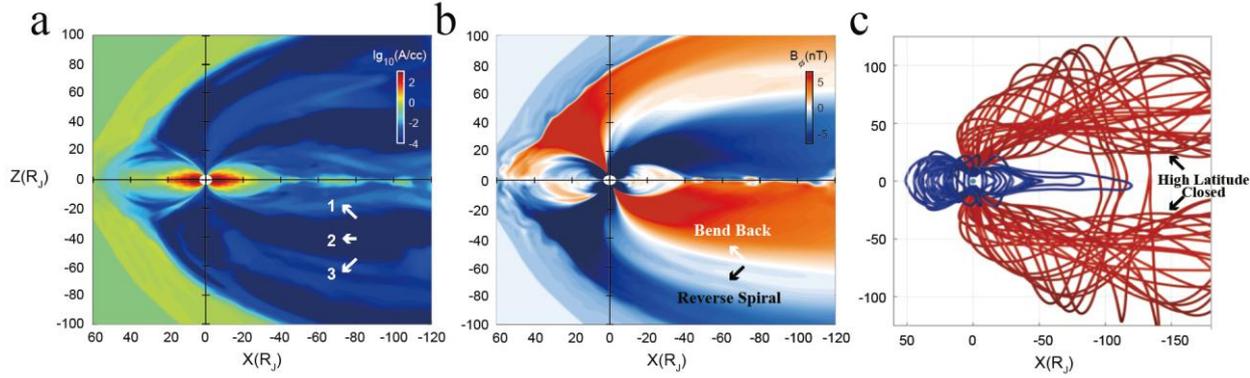

**Figure 4. High-latitude closed-flux topology with dense plasma and reversed azimuthal twist revealed by global MHD simulations.** Noon–midnight meridional cross-section from a global MHD simulation of Jupiter's magnetosphere in JSM (Jupiter Solar Magnetospheric) coordinates. (**a**) Plasma number density distribution. White arrows (1, 2, 3) indicate three regions: (1) the low-latitude current sheet with peak densities, (2) the mid-latitude lobe with depleted plasma, and (3) the high-latitude dense region ($|Z| > 50$ $R_J$) with elevated densities consistent with Juno observations (Fig. 1). (**b**) Azimuthal magnetic field component $B_\varphi$. The nightside high-latitude region exhibits a 'reverse spiral' structure, opposite in sense to the classical bend-back configuration of the equatorial magnetotail. This azimuthal reversal agrees with Juno measurements (Fig. 3c). (**c**) Three-dimensional magnetic field topology traced from the simulation. Blue field lines represent conventional low-latitude closed lines that exhibit the typical bend-back geometry below roughly 30 $R_J$. Red field lines represent high-latitude closed flux tubes originating from the polar regions ($|Z| > 40$ $R_J$) that extend into the distant magnetotail before returning planetward.

## Discussion

The identification of a persistent high-latitude dense plasma structure in Jupiter's nightside magnetosphere suggests that the canonical understanding of planetary magnetotail architecture need to be revised. The observed high-latitude structure provides direct observational evidence supporting theoretical predictions that Jupiter's polar caps are threaded predominantly by closed magnetic flux tubes extending deep into the magnetotail[5,41]. Its heavy-ion composition, including Io-sourced sulfur and oxygen, demonstrates a magnetospheric rather than solar-wind origin[17]. The similarity of ion populations in the HD region, the equatorial current sheet, and closed polar regions near Jupiter[24] suggests a continuous system of closed flux linking both polar caps through the outer magnetosphere, unlike Earth's open polar cap topology[17]. The fundamental difference arises from the competition between reconnection and rotation timescales: Jupiter's reconnection timescale substantially exceeds its rotation period[5]. This prevents the establishment of steady-state Dungey circulation that characterizes terrestrial magnetospheric convection[8,42], representing a different type of magnetospheric interaction with the solar wind[5,43,44].

The systematic reversal of azimuthal magnetic field orientation in the high-latitude dense region reveals the three-dimensional helical structure of closed polar flux tubes under rotational stress. While equatorial and mid-latitude regions exhibit the expected bend-back configuration with positive $B_\varphi$[45–47]—reflecting tailward stretching of field lines by outward plasma transport—the high-latitude structure displays negative $B_\varphi$, indicating field lines twisted in the opposite sense. This reverse spiral configuration arises from the differential rotation between the ionosphere and magnetospheric plasma: as closed field lines thread from one polar cap through the distant magnetosphere to the opposite hemisphere, they experience competing rotational forces that generate the observed helical topology. Global MHD simulations reproduce these field reversals and dense plasma features at high latitudes (Fig. 4).

In planetary magnetosphere, current sheets are typically associated with rapid magnetic field variation across a finite-thickness layer, such that the field gradients correspond to enhanced current density through Ampère's law[48]. At Jupiter, plasma and current are often concentrated in the same broad magnetodisc layer[16,46]. In this sense, the HD region shows the essential characteristics of a current-sheet-like layer: persistent sheet-like plasma enhancement, systematic azimuthal magnetic shear (Fig. 3c and 4b), and reduced radial field strength relative to the adjacent lobe (Fig. 3b). Although the reversal of $B_\varphi$ alone is not sufficient, combined with the layer geometry, plasma enhancement, and MHD support, it indicates that the HD region probably represents a high-latitude current-carrying layer on closed flux tubes.

This structure may help explain auroral generation and magnetosphere–ionosphere coupling at Jupiter[49,50]. Unlike Earth's polar cap aurora, which depends on solar wind particle precipitation along open field lines, Jupiter's polar emissions can arise from internal magnetospheric dynamics operating on closed flux tubes[20,50,51]. However, the specific processes governing the behavior of these high-latitude closed field lines remain unclear. Variability in Jupiter's polar emissions may provide a window into the dynamics of the HD regions[21,51,52]. The helical field configuration and associated current systems may provide a natural framework for understanding several key features of Jupiter's aurora, including transient polar structures and transpolar arcs that have long puzzled observers[18,21,53].

This study reveals dense plasma populations extending to high latitudes on closed flux tubes that connect both polar caps through Jupiter's nightside magnetosphere. Together with the reversed helical magnetic topology and heavy-ion composition, this indicates a rotation-dominated regime in which internal mass loading reshapes the magnetotail[15]. Future investigations can focus on quantifying the variability and current of the high-latitude structure associated with the entanglement of the magnetic field lines, quantifying energy transfer rates between the ionosphere and magnetosphere, and determining how this unique configuration influences magnetospheric dynamics during compression events. Understanding these processes is essential for predicting the magnetospheric environments of exoplanets and assessing their potential for maintaining atmospheres conducive to habitability.

## Methods

### Simulation Information

In this study, we use the three-dimensional global Jovian magnetosphere simulation described by Zhang et al.[5]. The calculations are performed with the Grid Agnostic MHD for Extended Research Applications (GAMERA) model[34], which solves the ideal magnetohydrodynamic (MHD) equations with a finite-volume approach on flexible curvilinear, non-orthogonal grids. This numerical framework is well suited for planetary magnetospheric applications because it can efficiently accommodate large-scale systems with strongly stretched and nonuniform geometries. The simulation is carried out in the solar-magnetic (SM) coordinate system. Its computational domain adopts a stretched spherical grid reaching 120 $R_J$ toward the Sun, -1300 $R_J$ down the magnetotail, and ±360 $R_J$ in the two directions transverse to the Sun–Jupiter line. Grid spacing is radially variable, becoming finer closer to the planet; near the inner boundary, the radial resolution is $\Delta r = 0.15$ $R_J$. The inner boundary itself is prescribed as a spherical surface located at 3.5 $R_J$ from Jupiter's center. Jupiter's internal field is approximated by a point dipole placed at the origin of the SM frame. To suppress hemispheric asymmetries and focus on the large-scale configuration, the dipole tilt is fixed at 0°. Rotational forcing is included through a steady corotation potential applied in the ionospheric potential formulation, following the magnetosphere–ionosphere coupling treatment of Merkin and Lyon[37] and Zhang et al.[54].

Plasma supplied by the Io torus is introduced through source terms added to the MHD equations, using the spatially dependent prescription given by Feng et al.[38]. In this study, the adopted Io mass loading rate is 1000 kg s$^{-1}$. The upstream driving conditions are held constant, with solar wind velocity components of $V_X = -380$ km s$^{-1}$, $V_Y = 0$, and $V_Z = 0$, and a solar wind number density of 0.2 cm$^{-3}$. The interplanetary magnetic field is specified as $B_X = 0$, $B_Y = 0.3$ nT, and $B_Z = 0$. These steady idealized inputs are used to examine the global topology of Jupiter's magnetosphere under controlled external forcing.

### Statistics Method for JADE TOF (time-of-flight) Data

To characterize the low-energy ion composition of the high-latitude dense region, we performed a statistical analysis of JADE time-of-flight (TOF) count rates. We first selected intervals that satisfied two criteria: magnetic latitude greater than 35° and a summed electron flux (100–1000 eV) exceeding 8000 counts s$^{-1}$. Adjacent intervals separated by less than 15 min were merged and treated as a single event. We then retained only events lasting longer than 5 min and containing more than five data points. Applying these criteria to all orbits included in this study yielded a total of 1123 events.

For each event, the TOF count rates were integrated over the full event duration. The mean event duration was 112 min, comparable to the ~90 min integration interval used for the closed polar flux tube TOF spectrum reported by Szalay et al. (*22*) and shown in Fig. 1d, thereby enabling a meaningful comparison. The statistical TOF spectrum shown in Fig. 1c was obtained by summing the TOF count rates from all selected events and dividing by the total number of events.

## Acknowledgement

The authors thank the Juno Magnetic Field Investigation (MAG) team for providing the high-quality magnetic field data used in this study.

## Data availability

All Juno data used in this study are publicly available from NASA's Planetary Data System (PDS) at https://pds-ppi.igpp.ucla.edu/. The magnetic field data are available from the Juno MAG calibrated data collection at https://pds-ppi.igpp.ucla.edu/collection/JNO-J-3-FGM-CAL-V1.0. The JADE plasma data are available from the Juno JADE calibrated data collection at https://pds-ppi.igpp.ucla.edu/collection/JNO-J_SW-JAD-3-CALIBRATED-V1.0. The simulation data for Jupiter's global density and magnetic field presented in this paper are publicly available at https://doi.org/10.17605/OSF.IO/8CQR4 .

## Code availability

The Juno data were processed and analyzed using the SPEDAS/PySPEDAS software package, which is publicly available at https://github.com/spedas. Derived data products and custom plotting scripts generated during this study are available from the corresponding author upon reasonable request. No new physical materials were generated in this study.

# Supplementary Information for
# Revealing a Systematic High-latitude Current Sheet at Jupiter

Y. Xu[1,2], Z. H. Yao[2]*, F. Allegrini[3,4], S. Y. Ye[1]*, B. Zhang[2]*, Z. L. Zeng[2], E. H. Feng[5], J. W. Sun[6], W. R. Dunn[7], S. J. Bolton[3]

[1] Department of Earth and Space Sciences, Southern University of Science and Technology (SUSTech), Shenzhen, China

[2] Department of Earth Sciences, the University of Hong Kong, Pokfulam, Hong Kong, SAR, China

[3] Southwest Research Institute, San Antonio, Texas, USA.

[4] Department of Physics and Astronomy, University of Texas at San Antonio, San Antonio, Texas, USA.

[5] School of Space and Earth Sciences, Beihang University, Beijing, China.

[6] School of Earth and Space Sciences, Peking University, Beijing, China.

[7] Mullard Space Science Laboratory, University College London, Dorking, UK.

Correspondence and requests for materials should be addressed to Zhonghua Yao (yaozh@hku.hk), Shengyi Ye (yesy@sustech.edu.cn) and Binzheng Zhang (binzh@hku.hk)

**Content**

Supplementary Figures 1-7

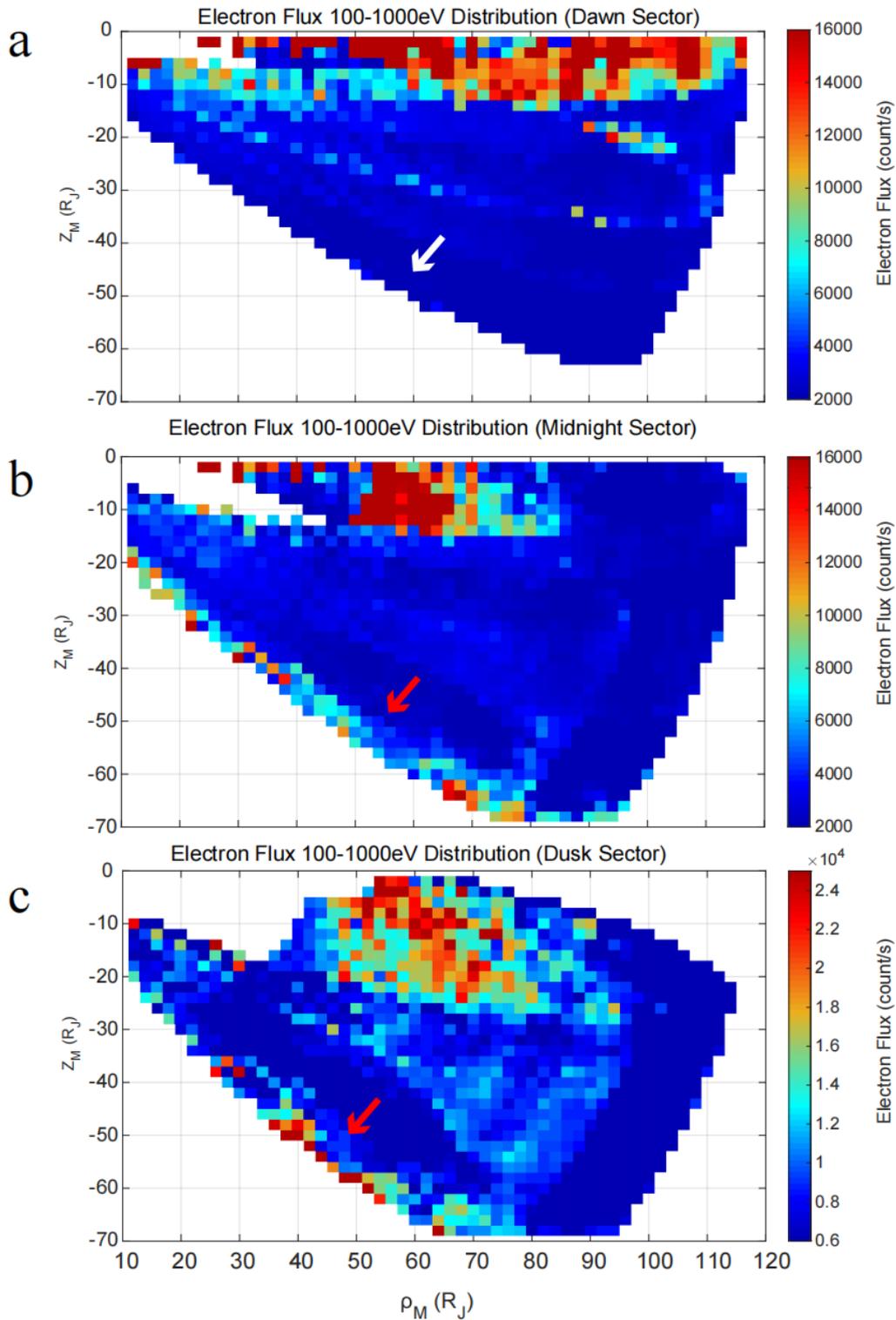

**Supplementary Fig. 1. Global distribution of total electron flux (100–1000 eV) across different local time sectors.** (**a**) Dawn/post-midnight sector, $0 \leq LT \leq 6$; (**b**) Midnight sector, $21 \leq LT \leq 3$; (**c**) Pre-midnight/dusk sector, $18 \leq LT \leq 24$. The white arrow highlights a low-latitude morning-side pass where the high-latitude dense (HD) region is not detected. The red arrow indicates that as Juno's orbital latitude increases in the midnight sector, the HD region becomes detectable.

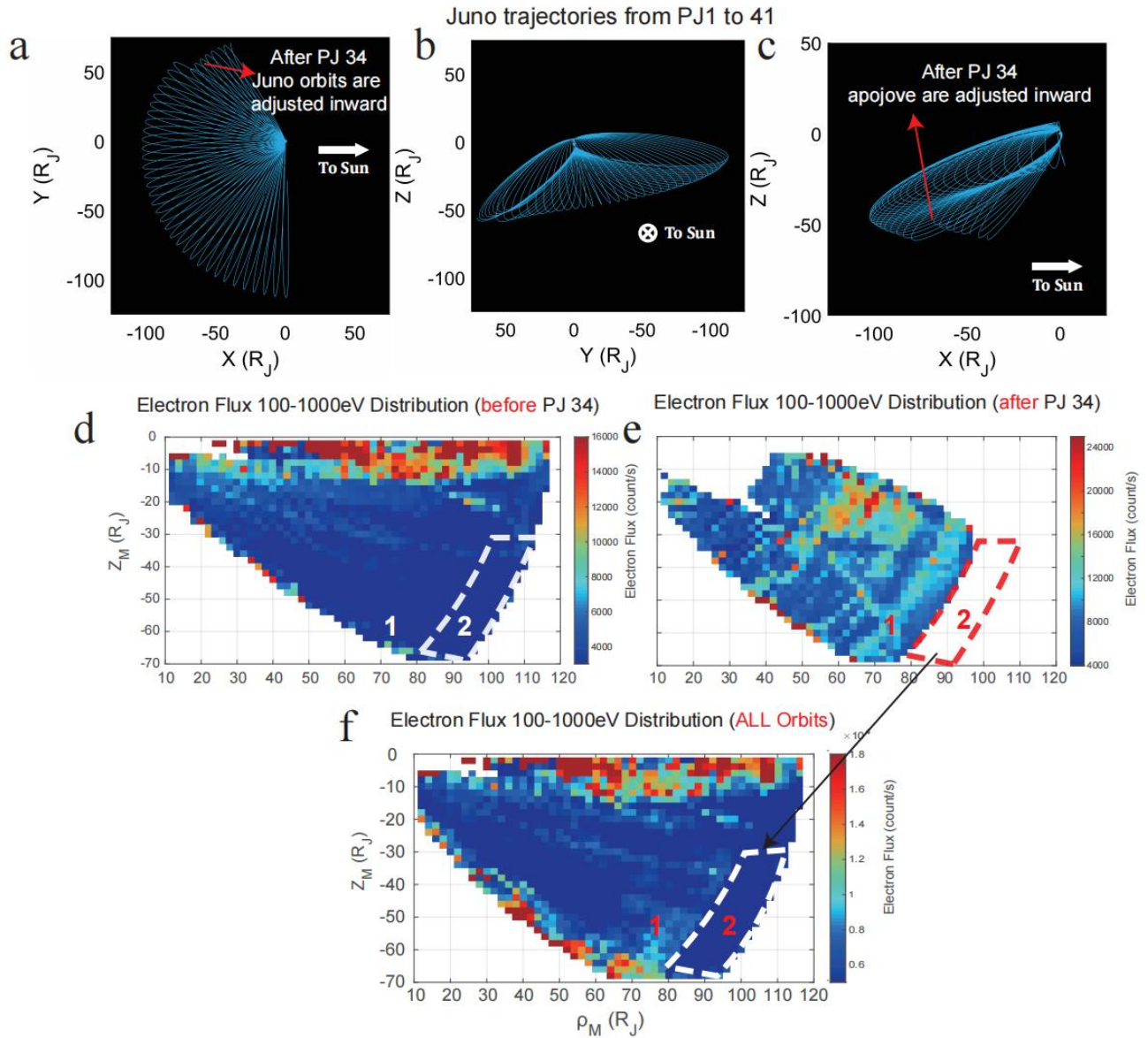

**Supplementary Fig. 2. Particle data drop caused by orbital effects of Juno spacecraft. (a-c)** Y–X, Z–Y, and Z–X views of the Juno spacecraft trajectory for Perijoves 1–41. After PJ34, Juno's orbit underwent a systematic inward contraction, which slightly affected the global mapping of particle data. (**d**) Global distribution of total electron flux (100–1000 eV) before PJ34. (**e**) Global distribution of total electron flux (100–1000 eV) after PJ34. (**f**) Global distribution of total electron flux (100–1000 eV) for all orbits combined. Because the dusk-side region exhibits relatively enhanced

particle activity, this orbital evolution introduces an apparent statistical decrease at approximately 15 $R_J$ from apojove in the all-orbit–averaged maps within region "2."

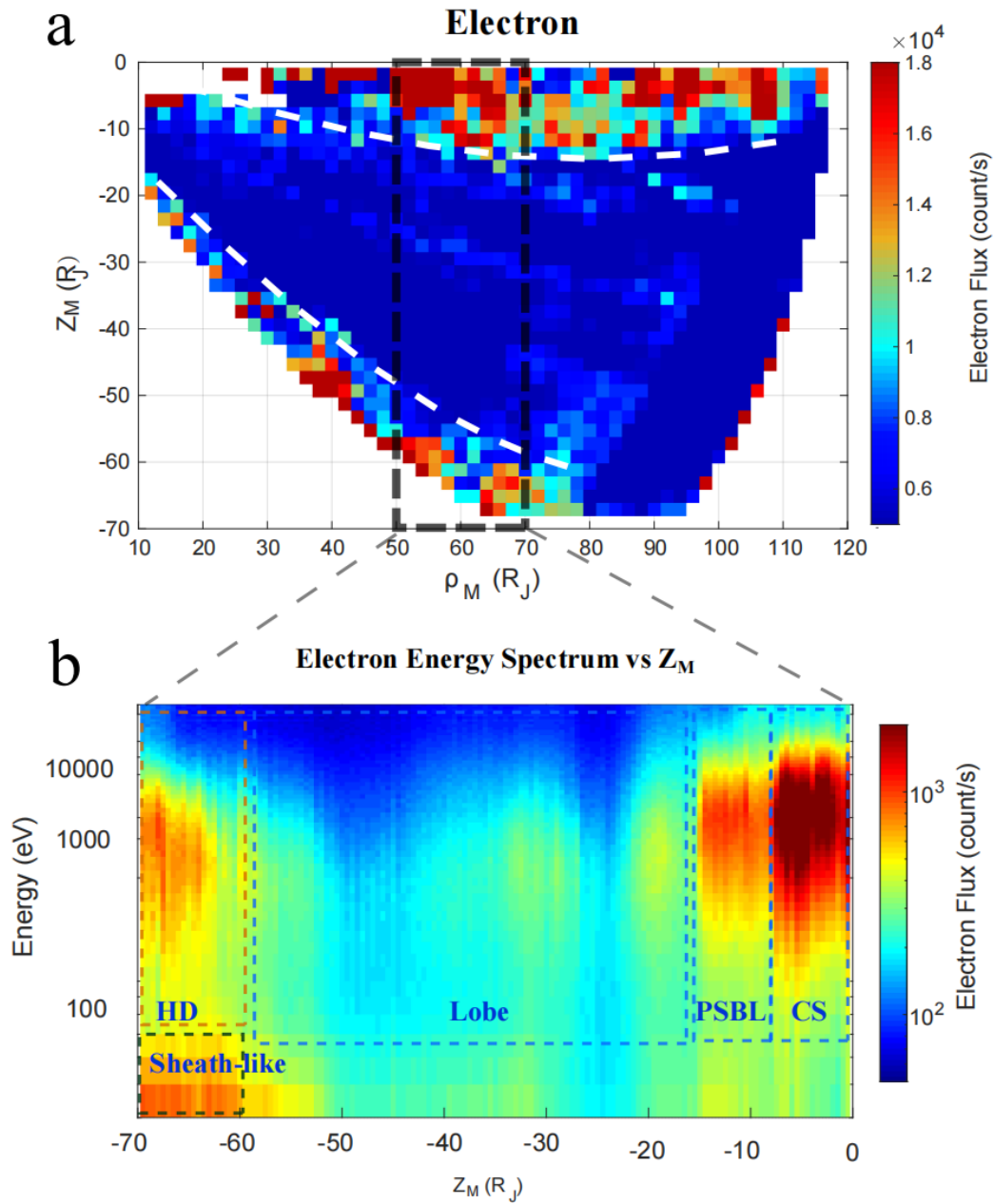

**Supplementary Fig. 3. Superposed epoch electron energy spectrum ($\rho_M$ range 50–70 $R_J$) at different heights from the magnetic equator.** (a) Statistical map of the total electron flux (100–1000 eV) in magnetic coordinates, reproduced from Fig. 1A; (b) Superposed epoch electron energy spectrum constructed from all electron measurements within $\rho_M$=50–70 $R_J$, showing how the electron energy distribution varies with magnetic height Z. This radial range was selected because the equatorial current sheet, the intervening relatively depleted lobe region, and the high-latitude HD region are most clearly separated in the vertical (Z) direction. With increasing

distance from the magnetic equator, the following regions can be identified: the plasma/current sheet (~−7 to 0 $R_J$), the plasma sheet boundary layer (~−15 to −7 $R_J$), the relatively empty lobe region (~−60 to −15 $R_J$), the HD region with current-sheet–like energy distributions, and the potential cusp region (23) with magnetosheath-like spectra (~−70 to −60 $R_J$).

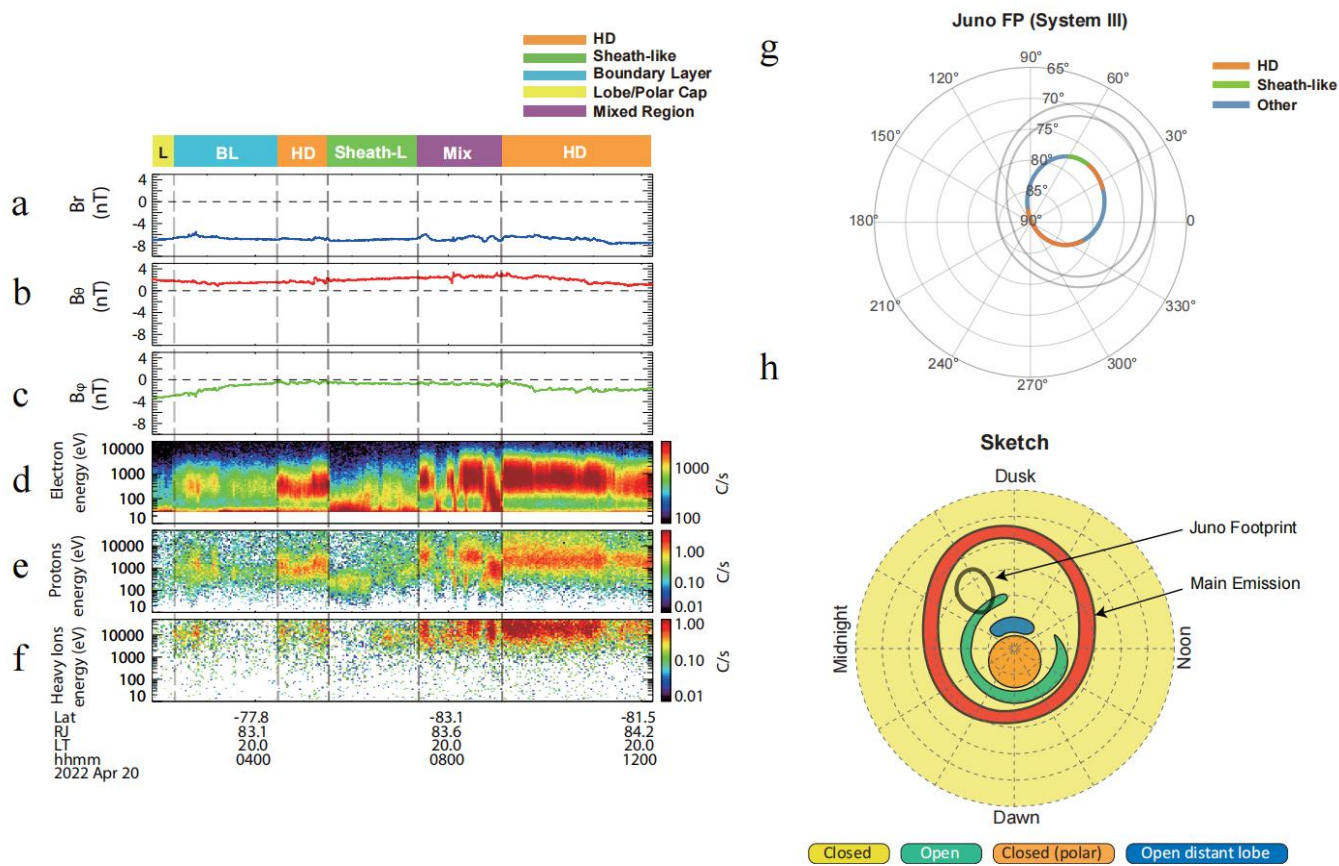

**Supplementary Fig. 4. High-latitude case observations in the pre-midnight sector on April 20, 2022. (a–c)** Magnetic field components ($B_r$, $B_\theta$, $B_\varphi$) in spherical magnetic coordinates from the Juno MAG instrument during a pre-midnight sector crossing. **(d–f)** Energy–time spectrograms of electrons, protons, and heavy ions from the JADE instrument, with color scales in counts per second. Colored bars at the top mark the identified plasma regions: polar cap (PC), boundary layer (BL), dense high-latitude current sheet (D), region with sheath-like signatures (Sheath-L), and mixed region. **(g)** Polar projection in System III coordinates showing the spacecraft's magnetic footpoint trajectory, mapped along JRM33 field lines[1] combined with the Connerney et al.[2] current sheet model. The black circle marks the main auroral oval. Color coding distinguishes the dense region (orange), region with sheath-like signatures (green), and other regions (blue, including BL, PC, and mixed regions). Throughout this interval, the footprints remained entirely poleward of the main auroral oval. **(h)** Footprint distributions corresponding to the open and closed

magnetic field regions in magnetic coordinates in the southern hemisphere based on the picture proposed by Zhang et al.[3]. In the pre-midnight high-latitude region, most spacecraft footpoints map to closed field lines (yellow area), with only occasional encounters with open field lines, during which cusp signatures may be observed.

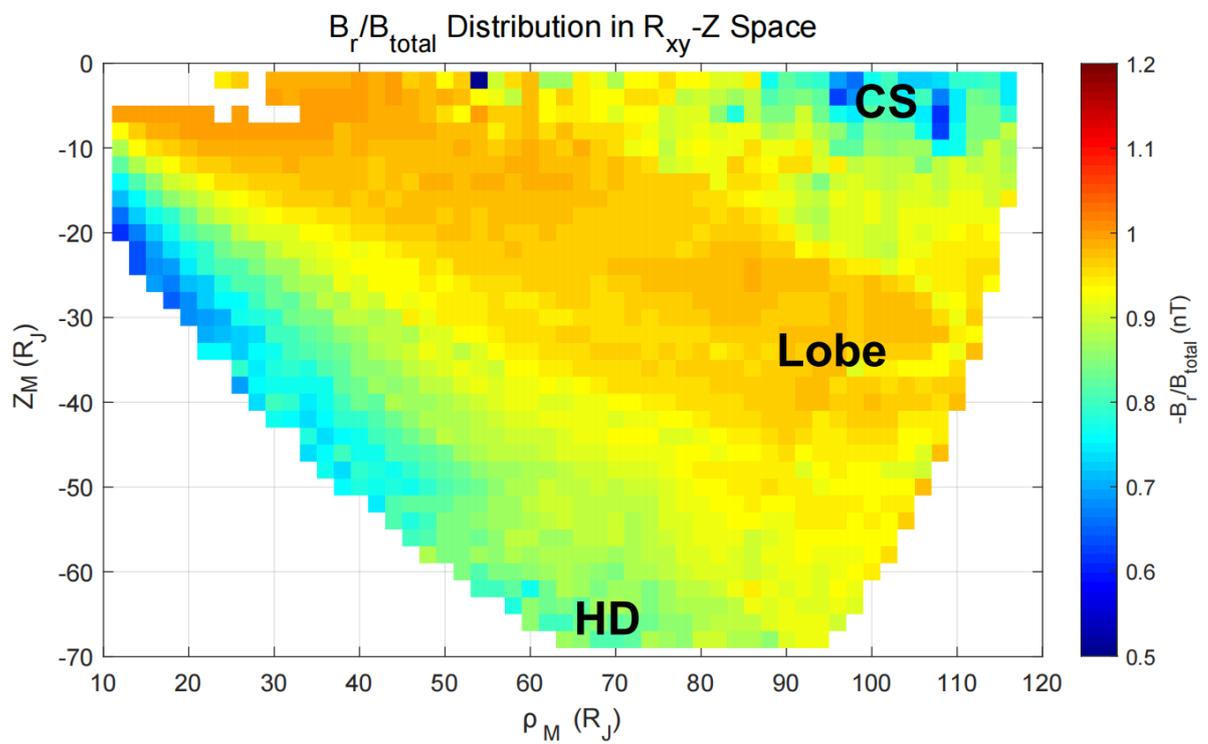

**Supplementary Fig. 5. The global distribution of |$B_r$|/$B_{total}$ corresponds to Fig. 3A in the main text.** This normalized representation more clearly highlights the systematic reduction of the radial magnetic field component in both the CS and HD regions.

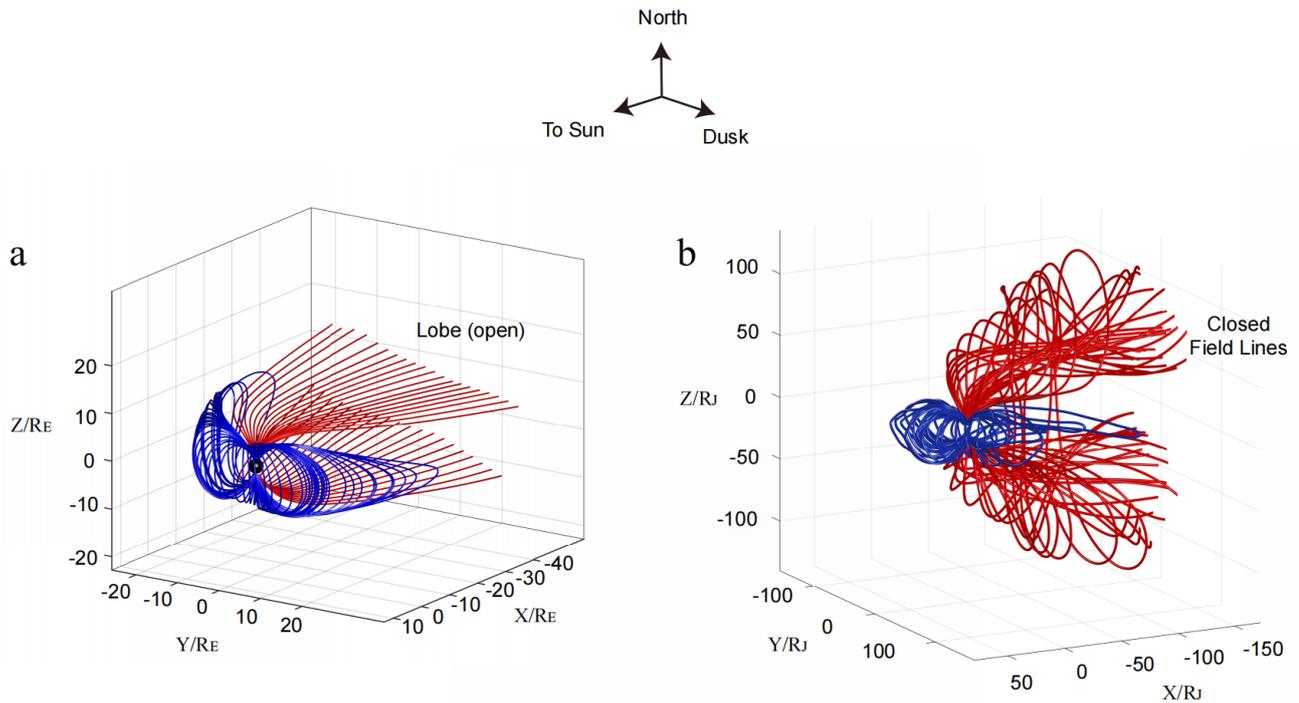

**Supplementary Fig. 6. Comparison of the high-latitude magnetic configurations in the simulations. (a)** Earth's open lobe structure, based on the results of Chen et al.[4]; **(b)** Jupiter's closed high-latitude field lines, from this study. Note that the red high-latitude field lines at Jupiter close in the distant magnetotail, as shown in Supplementary Fig. 7.

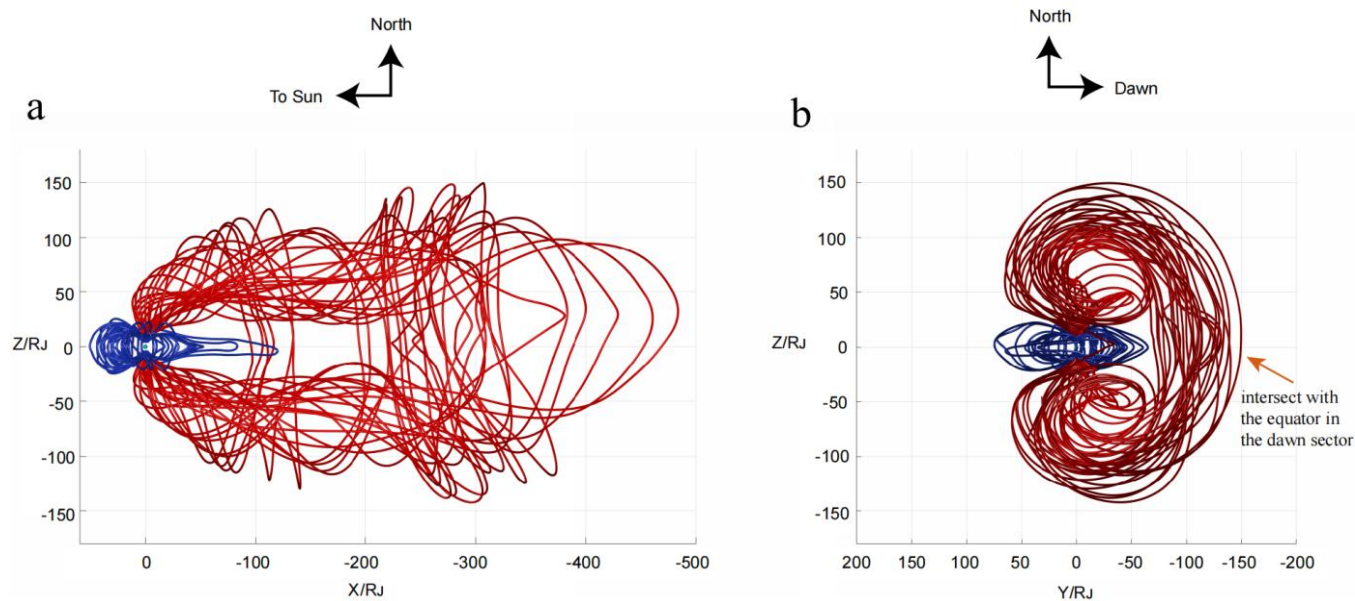

**Supplementary Fig. 7. Complete views of simulated Jupiter's high-latitude closed magnetic field lines in this study: (a)** X–Z view; **(b)** Y–Z view. The orange arrows indicate that the intersections of the closed field lines with the magnetic equator are all located on the dawn side, consistent with previous simulation results.

## Supplementary Materials References